\documentclass[preprint,aps,draft]{revtex4-1}

\usepackage{graphicx}% Include figure files
\usepackage{epsfig}
\usepackage{dcolumn}% Align table columns on decimal point
\usepackage{bm}% bold math
\usepackage[utf8]{inputenc}
\inputencoding{utf8}
\usepackage{multirow}
\usepackage{caption}
\usepackage{subcaption}
\usepackage{amssymb}
\usepackage{amsmath}
\usepackage{hyperref}

\usepackage{amsmath}%
\newcommand\trick[1]{}

\def\be{\begin{equation}}
\def\ee{\end{equation}}

\def\ba{\begin{eqnarray}}
\def\ea{\end{eqnarray}}
\usepackage{hyperref}

\begin{document}

\title{Stability Conditions for the Horndeski Scalar Field Gravity Model}

\author{Cláudio Gomes}
\email{E-mail: claudio.gomes@fc.up.pt}
\affiliation{Departamento de Ciências da Física, Química e Engenharia, Faculdade de Ci\^encias e Tecnologia da Universidade dos Açores, Campus de Ponta Delgada, Rua da Mãe de Deus 9500-321 Ponta Delgada, Portugal}
\affiliation{Centro de Física do Porto, Rua do Campo Alegre s/n, 4169-007 Porto, Portugal}

\author{Orfeu Bertolami}
\email{E-mail: orfeu.bertolami@fc.up.pt}
\affiliation{Departamento de F\'isica e Astronomia, Faculdade de Ci\^encias da Universidade do Porto, Rua do Campo Alegre s/n, 4169-007 Porto, Portugal}
\affiliation{Centro de Física do Porto, Rua do Campo Alegre s/n, 4169-007 Porto, Portugal}

\date{\today}

\begin{abstract}
We constrain the viable models of Horndeski gravity, written in its equivalent Generalised Galileon version, by resorting to the Witten positive energy theorem. We find that the free function $G_3(\phi,X)$ in the Lagrangian is constrained to be a function solely of the scalar field, $G_3(\phi)$, and relations among the free functions are found. Other criterion for stability are also analysed, such as the attractiveness of gravity, the Dolgov-Kawasacki instability and the energy conditions. Some applications for Cosmology are discussed.
\end{abstract}

\maketitle

%%%%%%%%%%%%%%%%%%%%%%%%%%%%%%%%%%%%%%%%%%%%%%%%%%%%%%%%%%%%%%%%%%%%%%%%%%%%%%%%%%%%%%
%%%%%%%%%%%%%%%%%%%%%%%%%%%%%%%%%%%%%%%%%%%%%%%%%%%%%%%%%%%%%%%%%%%%%%%%%%%%%%%%%%%%%%
%%%%%%%%%%%%%%%%%%%%%%%%%%%%%%%%%%%%%%%%%%%%%%%%%%%%%%%%%%%%%%%%%%%%%%%%%%%%%%%%%%%%%%
%%%%%%%%%%%%%%%%%%%%%%%%%%%%%%%%%%%%%%%%%%%%%%%%%%%%%%%%%%%%%%%%%%%%%%%%%%%%%%%%%%%%%%
%%%%%%%%%%%%%%%%%%%%%%%%%%%%%%%%%%%%%%%%%%%%%%%%%%%%%%%%%%%%%%%%%%%%%%%%%%%%%%%%%%%%%%

%\section{Introduction}
%\label{sec:intro}

General Relativity (GR) is the most well established theory of gravity to date. It leads to second order metric field equations, which are free from ghost instabilities, being compatible with several observational tests \cite{gr0,gr1}. However, in order to match astrophysical and cosmological data, two dark components are required, which comprise around $95\%$ of the energy content of the Universe and have not been directly detected so far. Furthermore, it lacks a consistent quantum version. Moreover, GR is not the most general theory which leads to second order field equations, hence higher dimensions theories such as Lovelock gravity \cite{lovelock}, higher order scalar curvature terms as f(R) theories \cite{fr} or non-minimally coupled curvature-matter theories \cite{nmc}, or scalar-tensor theories as Horndeski gravity constitute robust alternatives \cite{horndeski1,horndeski2}. In particular, Horndeski gravity is the most general extension of GR, in four-dimensional spacetime, involving a scalar field. Later on, Horndeski himself extended the model relying on a Abelian vector field whose action is invariant under $U(1)$ transformations, instead of the scalar field \cite{horndeskivector}. This model reduces in flat spacetime to the Einstein-Maxwell action \cite{einsteinmaxwell}, and if gauge invariance is relaxed one is left with the Proca action \cite{heisenberg,tasinato}. However the Horndeski scalar field work was ignored for decades until its was rediscovered in a different formulation, the Galileon action \cite{deffayet,generalisedgalileon,galileons}. A classical Galileon in flat spacetime is a field, $\pi$, which obeys to a Galilean symmetry, $\pi \to \pi +b_{\mu}x^{\mu}+c$, with $b_{\mu}$ and $c$ being a constant four-vector and a constant scalar, respectively. Its covariant generalisation breaks the Galilean symmetry, but gives origin to field equations of order non higher than two in the spacetime derivatives \cite{deffayet,generalisedgalileon,horndeskiformal}, in such way that both classical and quantum pathologies are absent. Hence, Horndeski scalar gravity and Generalised Galileon Gravity are equivalent to each other at least in four dimensions \cite{generalisedgalileon}. Moreover, Horndeski gravity encompasses GR, Brans-Dicke theory, Quintessence, Dilaton, Chameleon gravity models, or even $f(R)$ theories upon a suitable conformal transformation.

These scalar models have a rich lore of theoretical and cosmological implications (see, for instance, Refs. \cite{generalisedgalileon,horndeskiapplications}), as well as Horndeski's vector field model \cite{horndeskiapplications2,horndeskiapplications3}.

Moreover, recent gravitational data from the collision of two neutron stars \cite{gravitationalwavesneutronstars} imposes stringent restrictions on alternative gravity models \cite{baker,creminelli,sakstein,ezquiaga,gomesgws}, namely from the constraint on the speed of gravitational waves, $-3 \times 10^{-15} \leq c_g /c - 1 \leq 7 \times 10^{-16}$. For instance, two of the Horndeski free functions in the Lagrangian densities were shown to be restricted to $G_4=G_4(\phi)$ and $G_5\approx const.:= G_{5}^{(0)}$ \cite{ezquiaga}. Some constraints on the full Lagrangian also arise from an effective field approach on this and other modified gravity models \cite{defelice}. 

	However, even if classical and quantum pathologies can be avoided in the Horndeski scalar gravity, provided the action functional is not degenerated, the corresponding Hamiltonian may not be bounded from below \cite{rodriguez}. Moreover, an important criteria for gravitational theories is that they obey the positive energy theorem as shown in Refs. \cite{yau1,yau2} for GR, which admits an elegant proof as demonstrated by Witten \cite{witten}. In broad terms, this theorem states that the total gravitational Hamiltonian of an isolated system is nonnegative. Witten's work lead to improvements and generalisations of the positive energy theorem for non supersymmetric gravity theories \cite{nester,boucher,gibbons,bertolami1987,zarro}.
	
We report a further constraint on the viable Horndeski models in light of Witten's positive energy theorem. To this end, for technical reasons we shall employ Nester's version of Witten's work \cite{nester}.

%%%%%%%%%%%%%%%%%%%%%%%%%%%%%%%%%%%%%%%%%%%%%%%%%%%%%%%%%%%%%%%%%%%%%%%%%%%%%%%%%%%%%%%%
%%%%%%%%%%%%%%%%%%%%%%%%%%%%%%%%%%%%%%%%%%%%%%%%%%%%%%%%%%%%%%%%%%%%%%%%%%%%%%%%%%%%%%%%
%%%%%%%%%%%%%%%%%%%%%%%%%%%%%%%%%%%%%%%%%%%%%%%%%%%%%%%%%%%%%%%%%%%%%%%%%%%%%%%%%%%%%%%%

{\bf Horndeski gravity after GW170817} Taking into account the stringent constraints from gravitational data \cite{gravitationalwavesneutronstars}, the action functional of Horndeski gravity, written in the form of Generalised Galileon theories, reads:
\begin{equation}
S=\int d^4x \sqrt{-g} \left[ \sum_{i=2}^5 \mathcal{L}_i + \mathcal{L}_M\right]~, 
\end{equation}
where $g$ stands for the metric determinant, $\mathcal{L}_M$ is the matter Lagrangian density and $\mathcal{L}_i$ are the Horndeski Lagrangian densities defined as:
\begin{eqnarray}
&&\mathcal{L}_2 := G_2(\phi,X)~,\\
&&\mathcal{L}_3 := G_3(\phi,X)\square \phi~,\\
&&\mathcal{L}_4 := -G_4(\phi)R~,\\
&&\mathcal{L}_5 := G_{5}^{(0)}G_{\mu\nu}\nabla^{\mu}\nabla^{\nu}\phi ~,
\end{eqnarray}
where  $G_i(\phi,X)$ are arbitrary functions of the scalar field, $\phi$, and its kinetic term, $X:=\frac{1}{2}\nabla_{\mu}\phi\nabla^{\mu}\phi$, $R$ is the scalar curvature, and $G_{\mu\nu}$ is the Einstein tensor. Our choise for the metric signature is $(+---)$.

Varying the action with respect to the metric field yields \cite{generalisedgalileon}:
\vspace{-0.5cm}
\begin{equation}
\label{eqn:metricfieldequations}
G_{\mu\nu}=\hat{\kappa}\left[T_{\mu\nu}^M+\hat{T}_{\mu\nu}\right]~,
\end{equation}
where the effective gravitational constant is given by $\hat{\kappa}:=\frac{1}{2G_4(\phi)}$, and the effective energy-momentum tensor reads:
%\vspace{-0.5cm}
%\begin{widetext}
\begin{eqnarray}
\hat{T}^{\sigma}_{~\alpha}=&&\left[G_{2X}+G_{3X}\square\phi-2G_{3\phi}+2G_{4\phi\phi}\right]\nabla^{\sigma}\phi\nabla_{\alpha}\phi+\left[\delta^{\sigma}_{~\alpha}\left(-G_2-2G_{4\phi}\square\phi\right)+2G_{4\phi}\nabla^{\sigma}\nabla_{\alpha}\phi\right]+\nonumber\\
&&\left[-2G_{3X}\nabla^{(\sigma}X\nabla_{\alpha)}\phi+\delta^{\sigma}_{~\alpha}G_{3X}\nabla_{\lambda}X\nabla^{\lambda}\phi\right]+2\delta^{\sigma}_{~\alpha}\left(G_{3\phi}-2G_{4\phi\phi}\right)X~.
\end{eqnarray}
%\end{widetext}
%\vspace{-1cm}
where $G_{iX}:=\frac{\partial G_i}{\partial X}$, $G_{i\phi}:=\frac{\partial G_i}{\partial \phi}$, and $T_{\mu\nu}^M:=\frac{2}{\sqrt{-g}}\frac{\delta \left(\sqrt{-g}\mathcal{L}\right)}{\delta g^{\mu\nu}}$ is the energy-momentum tensor of matter.

{\bf Stability Criteria:} {\it Attractive Gravity} An important stability criterion concerns the avoidance of ghost modes and gradient instabilities, which translates into the demand on the positiveness of both numerator and denominator of the speed of sound of gravitational waves. For the Horndeski models, after GW170817, this implies that $G_4(\phi)>0$. This matches precisely the requirement of an attractive gravity model, $\hat{\kappa}>0$.

{\it Dolgov-Kawasacki instabilities} These instabilities \cite{dolgov} arise when a gravity theory allows for a dynamical equation for the scalar curvature through the trace of the metric field equations and the associated "square mass" is non-positive. Since the trace of the viable Horndeski models renders an algebraic equation for the scalar curvature, no Dolgov-Kawasacki occurs:
\begin{equation}
\label{eqn:trace}
R=-\frac{1}{2G_4}\left[T+\hat{T}\right]~,
\end{equation}
where $\hat{T}:=2G_{2X}X- 4G_2 +2G_{3X}\square\phi X +2\nabla_{\lambda}G_3\nabla^{\lambda}\phi-6\left[G_{4\phi}\square\phi+2G_{4\phi\phi}X\right]$.

{\it Witten positive energy theorem}  We start by relating the total energy-momentum tensor associated to a four-momentum vector $p^{\mu}$ for an asymptotically flat spacetime, given by the surface integral of the difference of two connections, $\Delta \Gamma^{\mu}_{\alpha\beta}=\Gamma^{\mu}_{\alpha\beta}- \Gamma^{\mu}_{\alpha\beta}(flat)=\mathcal{O}(1/r^2)$, with the integral of the two form  $E^{\sigma\alpha}:=2\left(\overline{\epsilon}\Gamma^{\sigma\alpha\beta}\nabla_{\beta}\epsilon-\overline{\nabla_{\beta}\epsilon}\Gamma^{\sigma\alpha\beta}\epsilon\right)$ given by \cite{witten}
\begin{equation}
\label{eqn:witten}
16\pi G V^{\mu}p_{\mu}=\frac{1}{2}\int_{S=\partial \Sigma}E^{\sigma\alpha}dS_{\sigma\alpha}=\int_{\Sigma}\nabla_{\alpha}E^{\sigma\alpha}d\Sigma_{\sigma}~,
\end{equation}
where $V^{\mu}:=\overline{\epsilon_0}\gamma^{\mu}\epsilon$, with $\epsilon_0$ being the value of the Dirac spinor $\epsilon$ at spatial infinity up to corrections of $\mathcal{O}(1/r)$, $\Sigma$ is an arbitrary 3-surface whose boundary at spatial infinity is $\partial\Sigma=S$, and we have used the following notations $\overline{\epsilon}=\epsilon^{\dagger}\gamma^0$, and $\gamma^{\mu}$ are the Dirac's $\gamma$ matrices. 

It can be shown that in a supersymmetric version of the positive energy theorem \cite{gibbons,boucher} the divergence of the previous two-form can be cast as:
\begin{equation}
\hat{\nabla}_{\alpha}E^{\sigma\alpha}= 2G^{\sigma}_{\alpha}\bar{\epsilon}^i\gamma^{\alpha}\epsilon^i+4\overline{\hat{\nabla}_{\alpha}\epsilon^i}\Gamma^{\sigma\alpha\beta}\hat{\nabla}_{\beta}\epsilon^i+\overline{\delta\chi^a}\gamma^{\sigma}\delta\chi^a~,
\end{equation}
where $\hat{\nabla}$ is the supersymmetric extension of the covariant derivative associated to the change of the gravitino field under a supersymmetric transformation,  $i=1,...,N$ is the total number of supersymetries, and $\delta\chi^a$ represents the change of a spin-$\frac{1}{2}$ field under supersymmetric transformation.

In GR, substituting the previous equation into Eq. (\ref{eqn:witten}), the positiveness of the integrand is ensured provided the energy-momentum tensor of matter fields satisfies the dominant energy condition, $\bar{\epsilon^i_0}\gamma^{\alpha}\epsilon_0^i$ is non-spacelike, and Witten's condition $\gamma^{\lambda}\hat{\nabla_{\lambda}}\epsilon^i=0$ is chosen. We note that the value of $\hat{\nabla_{\lambda}}\epsilon^i$ and $\delta\chi^a$ are set by supersymmetry \cite{gibbons,boucher}.

This procedure can be generalised to non-supersymmetric theories \cite{boucher, bertolami1987}, resulting in the introduction of three functions in the definitions of the gradient of the Dirac spinor and the change of the gravitino-like terms, to be later determined, in order to ensure the positiveness of the generalised version of the integrand of Eq. (\ref{eqn:witten}):
\begin{eqnarray}
\label{eqn:definitions}
&&\hat{\nabla_{\alpha}}\epsilon^i:=\nabla_{\alpha}\epsilon^i+\frac{i}{2}\hat{\kappa}\gamma_{\alpha}f_1^{ij}\epsilon^j~,\\
&&\delta\chi^a=i\gamma^{\lambda}\nabla_{\lambda}\phi f_2^{ai}\epsilon^i+f_3^{ai}\epsilon^i~.
\end{eqnarray}

In the case of Horndeski theories, the above free functions should depend on the scalar field and its kinetic term, i.e., $f_i=f_i(\phi,X), ~i=1,...,3$. Hence the divergence of the two-form $E^{\sigma\alpha}$ reads:
%\begin{widetext}
\begin{eqnarray}
\nabla_{\alpha}\hat{E}^{\sigma\alpha}&&=2\hat{\kappa}(\phi)T^{\sigma}_{\lambda}\bar{\epsilon}\gamma^{\lambda}\epsilon+4\nabla_{\alpha}\bar{\epsilon}^i\Gamma^{\sigma\alpha\beta}\nabla_{\beta}\epsilon^i+\overline{\delta\chi^a}\gamma^{\sigma}\delta\chi^a + \nonumber\\
&& + \left[-2f_2(\phi,X)^{ai}f_2(\phi,X)^{aj}+2\hat{\kappa}(\phi,X)\left(G_{2X}(\phi,X)+G_{3X}(\phi,X)\square\phi-\right.\right.\nonumber\\
&&\left.\left. -2G_{3\phi}(\phi,X)+2G_{4\phi\phi}(\phi)\right)\delta^{ij}\right]\nabla^{\sigma}\phi\nabla_{\alpha}\phi \bar{\epsilon}^i\gamma^{\alpha} \epsilon^i\nonumber\\
&&+ \left[ 6\delta^{\sigma}_{~\alpha}\left(\hat{\kappa}(\phi)f_1(\phi,X)^{ij}\right)^2-\delta^{\sigma}_{~\alpha}f_3(\phi,X)^{ai}f_3(\phi,X)^{aj}\right.\nonumber\\
&&\left.+2\hat{\kappa}(\phi)\left(-G_2(\phi,X)\delta^{\sigma}_{\alpha}-2G_{4\phi}(\phi)\square\phi\delta^{\sigma}_{\alpha}+2G_{4\phi}(\phi)\nabla^{\sigma}\nabla_{\alpha}\phi\right)\delta^{ij}\right]\bar{\epsilon}^i\gamma^{\alpha} \epsilon^j \nonumber\\
&&+ i \left[4\left(\hat{\kappa}(\phi)f_1(\phi,X)^{ij}\right)_{\phi}-f_2(\phi,X)^{ai}f_3(\phi,X)^{aj}-f_2(\phi,X)^{aj}f_3(\phi,X)^{ai}\right]\nabla_{\alpha}\phi\bar{\epsilon}^i[\gamma^{\sigma},\gamma_{\alpha}]\epsilon^j  \nonumber\\
&&+ 2i \left[f_2(\phi,X)^{ai}f_3(\phi,X)^{aj}-f_2(\phi,X)^{aj}f_3(\phi,X)^{ai}\right]\nabla^{\sigma}\phi\bar{\epsilon}^i\epsilon^j \nonumber\\
&&+ \left[-2f_2(\phi,X)^{ai}f_2(\phi,X)^{aj}+4\hat{\kappa}(\phi)\left(G_{3\phi}(\phi,X)-2G_{4\phi\phi}(\phi)\right)\delta^{ij}\right]X \bar{\epsilon}^i\gamma^{\sigma}\epsilon^j \nonumber\\
&&+ 2\hat{\kappa}(\phi)\left[-2G_{3X}(\phi,X)\nabla_{(\alpha}\phi\nabla^{\sigma)}X+\delta^{\sigma}_{\alpha}G_{3X}(\phi,X)\nabla^{\lambda}\phi\nabla_{\lambda}X\right] \bar{\epsilon}^i\gamma^{\sigma}\epsilon^j~.
\end{eqnarray}
%\end{widetext}

In order to identify with the general expression of Eq. (\ref{eqn:witten}), each expression for each extra term needs to vanish.

Thus, we can easily verify that the conditions arising: from the sixth line are always satisfied since $\bar{\epsilon}^i\epsilon^j=-\bar{\epsilon}^j\epsilon^i$; from the seventh line can be inserted in the second one, resulting a simpler condition, namely $G_{2X}+G_{3X}\square\phi-4G_{3\phi}+6G_{4\phi\phi}=0$; from the last line imply that $G_{3X}=0 \implies G_3=G_3(\phi)$.
 % This is a quite relevant result: the Witten positive energy theorem implies that $G_3$ can only be a function of $\phi$. This narrows down the viable models within Horndeski theory even further alongside with the recent gravitational data.
Thus, the set of equations becomes:
%\begin{widetext}
\begin{equation}
\label{eqn:finalset}
  \left\{\begin{aligned}
  G_{2X}(\phi,X)\delta^{ij}&=\left(4G_{3\phi}(\phi)-6G_{4\phi\phi}(\phi)\right)\delta^{ij}\\
  6\left(\hat{\kappa}(\phi)f_1(\phi,X)^{ij}\right)^2\delta^{\sigma}_{~\alpha}&=f_3(\phi,X)^{ai}f_3(\phi,X)^{aj}\delta^{\sigma}_{~\alpha}+\\
  &+2\hat{\kappa}(\phi)\left(G_2(\phi,X)\delta^{\sigma}_{\alpha}+2G_{4\phi}(\phi)\square\phi\delta^{\sigma}_{\alpha}-2G_{4\phi}(\phi)\nabla^{\sigma}\nabla_{\alpha}\phi\right)\delta^{ij}\\
  4\left(\hat{\kappa}(\phi)f_1(\phi,X)^{ij}\right)_{\phi}&=f_2(\phi,X)^{ai}f_3(\phi,X)^{aj}+f_2(\phi,X)^{aj}f_3(\phi,X)^{ai}\\
  f_2(\phi,X)^{ai}f_2(\phi,X)^{aj}&=2\hat{\kappa}(\phi)\left(-G_{3\phi}(\phi)+2G_{4\phi\phi}(\phi)\right)\delta^{ij}
\end{aligned}\right. ~, 
\end{equation}
%\end{widetext}
provided that $\hat{\kappa}(\phi)\neq 0$. This set also constrains the relations among the different functions $f_i(\phi,X)$ for each model with $G_2(\phi,X), ~G_3(\phi), ~G_4(\phi), ~G_{5}^{(0)}$.

By comparing our results with the $N=1$ Supergravity, where $f_3=0$, \cite{boucher,bertolami1987} a boundary condition is found:
\begin{equation}
f_1^{ij}(\phi_0)=\sqrt{\frac{G_2(\phi_0)}{6G_4(\phi_0)}}\delta^{ij}~,
\end{equation}
where $\phi_0$ is the value at spatial infinity of the stationary point of the potential encoded in the function $G_2$. 
%In fact, the effective potential will be of the form of the quotient $\bar{G}_2/G_4$, where $\bar{G}_2$ is the potential part of the function $G_2$, as it will be intuitive further ahead.

Furthermore, we can state that such configuration is stable in the sense of the Witten's theorem and Refs. \cite{witten, boucher, bertolami1987,zarro}, provided the set of Eqs. (\ref{eqn:finalset}) can be solved together with the above boundary condition.

{\it Scalar field equations} Varying the action functional relatively to the scalar field, and taking into consideration both gravitational waves data and our results, we get:
\begin{eqnarray}
&&G_{2\phi}-2G_{2X\phi}X-G_{2XX}\nabla^{\mu}X\nabla_{\mu}\phi-G_{2X}\square\phi+\nonumber\\
&&+2G_{3\phi}\square\phi+2G_{3\phi\phi}X-G_{4\phi}R=0~.
\end{eqnarray}

Solving this equation relatively to the Ricci scalar, and inserting into Eq. (\ref{eqn:trace}), we get a further relation between the free functions in the Lagrangian densities of Horndeski theory:
\begin{eqnarray}
&&G_{2\phi}-2G_{2X\phi}X-G_{2XX}\nabla^{\mu}X\nabla_{\mu}\phi-G_{2X}\square\phi+2G_{3\phi}\square\phi+\nonumber\\
&&+2G_{3\phi\phi}X=-\frac{1}{2}T-G_{2X}X+2G_2-2G_{3\phi}X+3G_{4\phi}\square\phi+\nonumber\\
&&+6G_{4\phi\phi}X~.
\end{eqnarray}

{\it Zero energy states} A comparison with the work of Ref. \cite{boucher}, a state of zero energy corresponds to the case where, in addition to obeying Witten's condition and $\delta \chi^a$ for all $\epsilon^i$, we have both $f_3=0$ and either $f_2=0$ or $\phi=\phi_0$. This leads to the following solutions:
\begin{eqnarray}
&&f_3=0~,\\
&&f_1=f_1(\phi)=(c_1\phi+c_2)2G_4(\phi)=\pm\sqrt{\frac{G_2}{6G_4}}~,\\
&&G_{3\phi}=\frac{3}{2}G_{4\phi\phi}~,\\
&&f_2^2=f_2^2(\phi)=\frac{G_{3\phi}}{3G_4}~.
\end{eqnarray}
 
Hence, zero energy states are found to be stable for $f_3=0$ and $\phi=\phi_0$, provided we can solve the above system of equations (once one of the $G_i$ functions are given, the system can, in principle, be fully solved).

A further criterion concerns the Strong, Null, Weak and Dominant Energy conditions, as we shall explore in the next subsection.

{\it Energy Conditions} For the discussion on the Energy conditions for the viable models of Horndeski, we shall closely follow the analysis performed in Ref. \cite{sequeira} for the non-minimal matter-curvature coupling alternative gravity model.

The Strong and the Null Energy Conditions arise from the purely geometric, and thus model independent, Raychaudhury equation together with the requirement of attractive gravity \cite{hawking}. It states that time variation of the expansion, $\theta$, of a congruence given by a vector field, $u^{\mu}$, for timelike geodesics reads:
\begin{equation}
\frac{d\theta}{d \tau}=-\frac{1}{3}\theta^2-\sigma_{\mu\nu}\sigma^{\mu\nu}+\omega_{\mu\nu}\omega^{\mu\nu}-R_{\mu\nu}u^{\mu}u^{\nu}~,
\end{equation}
where $\sigma_{\mu\nu}$ is the shear and $\omega_{\mu\nu}$ is the rotation tensors associated with the congruence. As for the case of null geodesics, the above expression is modified for a  future-pointing null vector field, $k^{\mu}$, as:
\begin{equation}
\frac{d\theta}{d \tau}=-\frac{1}{2}\theta^2-\sigma_{\mu\nu}\sigma^{\mu\nu}+\omega_{\mu\nu}\omega^{\mu\nu}-R_{\mu\nu}k^{\mu}k^{\nu}~.
\end{equation}

For any hypersurface of orthogonal congruences, where $\omega_{\mu\nu}=0$, and since $\sigma_{\mu\nu}\sigma^{\mu\nu}>0$, the requirement of attractive gravity, $d\theta/d\tau<0$, translates into:
\begin{eqnarray}
&&R_{\mu\nu}u^{\mu}u^{\nu}\geq 0~,\\
&&R_{\mu\nu}k^{\mu}k^{\nu}\geq 0~,
\end{eqnarray}
where the first condition is known as the Strong Energy Condition (SEC), and the second one as the Null Energy Condition (NEC). It is straightforward to see that the metric field conditions, Eqs. (\ref{eqn:metricfieldequations}), can be used in order to impose the above conditions for the Horndeski viable models. To do so, let us note that the Ricci scalar in the metric field equations can be substituted by the equality arisen from the trace of those. Hence, we get for the SEC:
\begin{equation}
\hat{\kappa}(\phi)\left[T_{\mu\nu}+\hat{T}_{\mu\nu}\right]u^{\mu}u^{\nu}-\frac{1}{2}\hat{\kappa}(\phi)\left[T+\hat{T}\right]\geq 0~,
\end{equation}
whilst for the NEC, we obtain:
\begin{equation}
\hat{\kappa}(\phi)\left[T_{\mu\nu}+\hat{T}_{\mu\nu}\right]k^{\mu}k^{\nu}\geq 0~,
\end{equation}

The remaining energy conditions, Weak (WEC) and Dominant Energy Conditions (DEC), concern the components of the energy momentum tensor. In particular, the WEC states that for every timelike vector field, $v^{\mu}$,  the matter density as measured by the corresponding observer is:
\vspace{-0.5cm}
\begin{equation}
\rho = T_{\mu\nu}v^{\mu}v^{\nu} \geq 0~,
\end{equation}
whilst the DEC states that for every future-pointing causal timelike or null vector field, $X^{\mu}$, the vector field, $T^{\mu}_{~\nu}X^{\nu}$, is a future-pointing causal vector field, i.e., for a perfect fluid:
\begin{equation}
\rho-p \geq 0. 
\end{equation}

In the case of the viable Horndeski models, these conditions can be straightforwardly generalised:
\begin{eqnarray}
\rho + \hat{\rho} \geq 0~,\\
\rho+\hat \rho - p- \hat{p}\geq 0~,
\end{eqnarray}
where $\hat{\rho}:=\hat{T}^0_0$ and $\hat{p}=-\frac{1}{3}\hat{T}^i_i$.

{\bf Cosmological implications} For a homogeneous and isotropic Universe, the line element is given by the Robertson-Walker metric:
\begin{equation}
ds^2=dt^2- a^2(t)dx^idx_i ~,
\end{equation}
$a(t)$ being the scale factor and we are considering a spacetime with no spatial curvature, as data suggest.

Consistently, the scalar field should be homogeneous and isotropic, $\phi=\phi(t)$. This allows for an identification of the energy-momentum tensor for the scalar field with a perfect fluid energy-momentum tensor:
\begin{eqnarray}
\label{eqn:perfectfluid}
\hat{\rho}:=&&\hat{T}^0_0=(G_{2X}-G_{3\phi})\dot{\phi}^2-G_2~,\\
\hat{p}:=&&-\frac{1}{3}\hat{T}^i_i=-\left[(-G_2-2G_{2\phi}\ddot{\phi})+ (G_{3\phi}-2G_{4\phi\phi})\dot{\phi}^2\right]~.
\end{eqnarray}

{\it Inflation} The hot Big Bang model, in order to be fully consistent with data, requires inflation to account for the homogeneity, isotropy, spatial flatness of the Universe, and the absence of topological defects such as magnetic monopoles. The latter also provides the seeds for large scale structure formation. Several models of inflation are consistent with data with impressive precision \cite{planck,exoticinflation,nmcinflation}.

A common setup for inflation rely on a scalar field slow-rolling down its potential. During this phase, the potential energy dominates the kinetic term, and for the viable Horndeski models, we can see from Eq. (\ref{eqn:perfectfluid}) that $\hat{\rho}\approx-G_2=-\hat{p}$. If we write $G_2(\phi,X)=X-V\approx-V$, we retrieve the usual description of scalar field inflation in GR.

Furthermore, the Friedmann equation, obtained from the time-time component of the metric field equations, becomes:
\vspace{-0.5cm}
\begin{equation}
3H^2=\frac{1}{2G_4(\phi)}\left[\rho_m+\hat{\rho}\right]~,
\end{equation}
which is of the standard form $H^2\approx \frac{8\pi G}{3} V(\phi)$ for $G_4(\phi)\neq 0$ and in the absence of other matter fields.

This means that during slow-roll inflation, Horndeski viable models should behave as the standard GR case although with $V(\phi)^{eff}\sim G_2(\phi)/G_4(\phi)$.

Another interesting cosmological scenario is the inclusion of a dominating cosmological constant, as it will be seen in the next subsection.

{\it Cosmological Constant} Comparing the definition of Eq. (\ref{eqn:definitions}) with $N=1$ Supergravity allows for the identification $\frac{1}{2}\kappa (\phi)f_1(\phi)^{ij}=\left(\pm\frac{1}{12}\Lambda\right)^{1/2}$ \cite{boucher}, where the $+$ sign occurs when $\Lambda>0$, and the $-$ sign when $\Lambda<0$ to ensure that the energy is positive definite. In addition, one has $f_3=0$.

For simplicity, we can assume that $i,j$ are single valued. This leads to
\begin{eqnarray}
&&f_1=f_1(\phi)=2\sqrt{\frac{\Lambda}{3}}G_4(\phi)~,\\
&&G_2=G_2(\phi)=2\Lambda G_4(\phi)~,\\
&&G_3(\phi)=\frac{3}{2}G_{4\phi\phi}~,\\
&&f_2=f_2(\phi)=\frac{G_{3\phi}}{3G_4}~.
\end{eqnarray}

Hence, once again the quotient $G_2(\phi)/G_4(\phi)$ appears related to the effective potential, in this case leading to the cosmological constant. In this sense, the cosmological constant is the quotient of two cosmological functions of the Lagrangian.

{\it Energy Conditions} The previous general energy conditions considerations can be applied for a homogeneous and isotropic Universe, where matter fields are assumed to be well described by a perfect fluid. For the SEC, we obtain:
\begin{equation}
\rho + 3p + 2G_2 +6G_{2\phi}\ddot{\phi}\geq 0~,
\end{equation}
where we have used the condition $(G_{2X}-4G_{3\phi}+6G_{4\phi\phi})\dot{\phi}^2=0$ from Witten's theorem.

As for the NEC, we obtain:
\begin{equation}
\rho+p+\left(G_{2X}-2G_{3\phi}+2G_{4\phi\phi}\right)\dot{\phi}^2+2G_{2\phi}\ddot{\phi}\geq 0~,
\end{equation}
and for the WEC, it is easy to show that:
\begin{equation}
\rho + (G_{2X}-G_{3\phi})\dot{\phi}^2-G_2 \geq 0~.
\end{equation}

Finally, as far as the DEC is concerned, we find:
\begin{equation}
\rho-p+\left(G_{2X}-2G_{4\phi\phi}\right)\dot{\phi}^2-2G_2-3G_{2\phi}\ddot{\phi}\geq 0.
\end{equation}

We point out that for accelerated expansions epochs, such as inflation, or the late time expansion motivated by a dark energy-like mechanism, the Strong Energy Condition is violated.

{\bf Conclusions} We have shown that, in light of Witten's theorem, the only viable Horndeski models are the ones which, in addition to GW data, have $G_3=G_3(\phi)$ together with a set of constraints among the remaining functions.

In a cosmological context, for instance, the ensemble of these constraints can be selective enough to allow for a quite small set of viable models.

C.G. is supported by the Fundo Regional para a Ciência e Tecnologia and Azores Government Grant No. M3.2DOCPROF/F/008/2020.

\end{document}